# The field-standardized average impact of national research systems compared to world average: the case of Italy[1]

Giovanni Abramo[a,b,*], Ciriaco Andrea D'Angelo[b], Fulvio Viel[b]

[a] *National Research Council of Italy*

[b] *Laboratory for Studies of Research and Technology Transfer, Department of Management, School of Engineering, University of Rome "Tor Vergata" - Italy*

**Abstract**

The study presents a time-series analysis of field-standardized average impact of Italian research compared to the world average. The approach is purely bibliometric, based on census of the full scientific production from all Italian public research organizations active in 2001-2006 (hard sciences only). The analysis is conducted both at sectorial level (aggregated, by scientific discipline and for single fields within disciplines) and at organizational level (by type of organization and for single organizations). The essence of the methodology should be replicable in all other national contexts. Its offers support to policy-makers and administrators for strategic analysis aimed at identifying strengths and weaknesses of national research systems and institutions.

# 1. Introduction

Under today's knowledge-based economy, governments of industrialized nations face pressing demands, particularly for ever-more effective scientific infrastructure, in order to support the competitiveness of their entire economic systems. These countries are turning to national research evaluation exercises, to pursue all or parts of the following objectives: i) stimulus of greater efficiency in research activity; ii) resource allocation based on merit; iii) demonstration that investment in research is effective and delivers public benefits; and iv) reduction of information asymmetry between supply of new knowledge and demand.

As the culture of evaluation extends, researchers, institutional administrators and policy-makers all feel the obligation and need to compare to others, both at the national and the international level. Comparative international-level analysis has potential strategic uses, for example for showing strengths and weaknesses in domestic research infrastructure, or for identifying capacities and rising opportunities to play leading roles in global growth of technological-scientific knowledge. Analysis could also show centers of excellence, by discipline and administrative region, with indications for research and industrial policy.

Scholars and practitioners have perceived the needs for comparative international evaluation and have attempted to provide answers, but the tasks involved are exceptionally complex. Sadly, the "results" we see have often been traded off for a loss of scientific rigor.

Various supranational organization (OECD, Eurostat, etc.) provide annual statistics on inputs to national research systems: gross expenditures on R&D (GERD), ratios of GERD to gross national product, number of researchers, etc. These inputs are also indicated by source of funds (public and private) and sector of performance (higher education, research organizations, business enterprise). The same organizations also offer statistics on outputs, such as number of publications, citations, patents, etc. Given this convenient data, there is a strong temptation to simply divide aggregate output values by aggregate input values, and thus provide handy international comparative measures of scientific productivity (Abramo and D'Angelo, 2007). As appealing as this procedure seems, it is fundamentally and seriously flawed. By now, all informed practitioners should most certainly be aware of the varying intensity of publications and citations among scientific disciplines (Moed, 2005; Radicchi et al., 2008). Thus, there should no longer be any thought of conducting performance comparisons without the necessity of applying field-standardization. Such studies, if ever produced, should never be published or publicized. The ritual brief warning on "interpretation and use of findings" is an exercise in false comfort. The scientific method imposes that the degree of accuracy of the measure be indicated in precise terms, for any type of analysis, and not with generic wording. Abramo et al. (2008) have in fact shown an order of magnitude for non-accuracy in comparative performance measures, when done without field-standardization. They compared performance rankings under aggregate and field-standardized measurement for all Italian universities active in the hard sciences: the comparison showed substantial variations in the rankings, with different placements for almost all the universities in every discipline. The problem of accuracy in productivity assessments is heavily related to issues of data aggregation. Comparisons at the gross aggregate level can never be taken in serious consideration, due to the differences in prolificacy of the research disciplines, and the differences in representativeness of the journals covered in source databases. Yet aggregate-type studies have even been published, especially at the outset, in journals of high impact and notable prestige as seen in articles for *Science* (May, 1997) and *Nature* (King, 2004), and also in specialized bibliometrics journals, such as *Scientometrics* (Braun et al., 1995). The first notable work on the subject was actually the one by Braun et al. (1995). The authors carried out comparative analysis of scientific production in 27 disciplines for the years 1989-1993, using the ISI Science Citation Index as their data source to compare the 50 nations with at least 1000 publications over the period. They conducted the comparison by applying the "Mean Expected Citation Rate", which was simply the ratio between citations for a publication and the impact factor of the journal in which it was published. In another example, May (1997) compared scientific research outputs among 15 leading countries. The dataset was the ISI-Science Citation Index for more than 4,000 journals, over the 14 years of 1981 to 1994. The analysis was simply based on the ratio between world share of publications and their relative citations. GDP spent on R&D and population size were used for standardizing total publications and total citations and comparing countries. In 2004, King updated and extended the work done by May, using very similar metrics to indicate comparative performance in science and engineering for a range of nations. The author analyzed the numbers of publications and their citations, provided by Thomson Scientific, in more than 8,000 journals. King noted a few cautions, particularly that citation (and, we would add, publication) analyses should not be used to compare different disciplines. Nevertheless, King himself cannot resist the temptation of providing international rankings of aggregate scientific performance (publications, citations, publications per researcher, citations per researcher, etc.). All of these are distorted by the lack of accounting for the relevant scientific specialization. The data in these studies were next taken up in a number of publications on research policy, adding a more serious side to the weaknesses in approach. For example, Dosi et al. (2006), take King's findings to demonstrate the inexistence of the "European Paradox"[2], and King's approach was next updated by Leydesdorff et al. (2009) to determine if the "United States are losing ground in science".

Another trend is the publication of international rankings of individual research organizations. Examples are the annual *Academic Ranking of World Universities (ARWU)*[3], provided by Shanghai Jiao Tong University, the *Times Higher Education World University Rankings*[4], and the *Performance Ranking of Scientific Publications for World Universities*, by the Higher Education Evaluation and Accreditation Council of Taiwan[5], etc. In these classifications, the performance indicators are given different weight in determining the position of universities. However, their use presents distortions both due to the lack of field-standardization and to strong size-dependency. The ARWU, for example, is notorious for the fact that over 90% of the performance result depends on university size. These non-scientific exercises are given more coverage in popular and promotional media and less in the scientific press.

International comparison is essential in all spheres, and particularly in scientific research, but the authors argue that the highly desirable ends clearly do not justify neglect of the scientific method. Given current capabilities in the field, the true comparison of the productivity of nations is still a distant goal. Our proposal is more modest and less appealing, but it is definitely one that permits accuracy: to compare the average impact value of national scientific production when standardized by scientific field. The authors wish to immediately emphasize that the comparison deals only with average impact of national publications, and that an average impact above the world average, even though field-standardized, still does not necessarily indicate superior productivity, because we do not know the number of publications per researcher in each field.[6] Abramo et al. (2010) have demonstrated, at least for the Italian case, that there is a strong correlation between quantity and quality of research production of individual scientists. But even if this were the general case, it still would not permit the conclusion that greater average impact per publication directly corresponds to greater total impact per researcher.

---

[2] This refers to "the conjecture that EU countries play a leading global role in terms of top-level scientific output, but lag behind in the ability of converting this strength into wealth-generating innovations" (Dosi et al., 2006).
[3] http://www.arwu.org/ (last accessed 18/02/2011)
[4] http://www.timeshighereducation.co.uk/world-university-rankings/ (last accessed 18/02/2011)
[5] http://ranking.heeact.edu.tw/en-us/2010/homepage/ (last accessed 18/02/2011)

Our literature search reveals only two extensive time-series analysis of the field-standardized average impact of national research systems that would be similar to the one presented here. The first concerns the international share of publications and citations for the Chinese science system, based on 35,000 publications extracted from the Science Citation Index for the period 1987-1993. The second is from Ingwersen (2009) and is about the development of research in Brazil compared to Mexico, Republic of South Africa (RSA) and the world, in two five-year periods 1996-2005.

We must mention also two Dutch studies of primarily methodological character, validated under very limited field of observation. Moed et al. (1985) provide a study of international impact for the faculties of medicine, natural sciences and mathematics at the University of Leiden. Van Leeuwen et al. (2003) test a proposed assessment method with a dataset of 18,000 publications in chemistry and related fields authored in 1991-2000, by 600 scientists of ten Dutch universities.

The present study takes a purely bibliometric approach, first mapping the scientific output of the Italian research system, according to the average impact achieved per field and organization. The objective is to compare the average impact of scientific production, for all Italian universities and public research organizations (U&PROs) active in the hard sciences, to the world average. In Italy there are 345 such U&PROs active in the hard sciences. The field-standardized analysis covers the period 2001-2006 and is presented at various sectorial levels (aggregate, inter-temporal, by discipline, by single field) and by organization (single organization and aggregated by typology of organization).

The study is relevant to various types of stakeholders. The findings should be useful in shaping policy interventions to consolidate excellence and reinforce weak fields that are considered strategic for national development. The inter-temporal analysis can provide useful in examining the effectiveness of interventions. For U&PRO administrators the study provides a benchmarking system that could be used in strategies to develop the institution's internal disciplines. The methodology, while applied to Italy, should essentially be replicable in all other national contexts.

The next section of the publications details the methodology used for the study. Section 3 presents the results from the analyses conducted, divided in three subsections: 3.1 presents the results of the aggregate national analysis and trends observed; 3.2 and 3.3 present, respectively, the analysis at the level of scientific discipline and field and at the level of organizations. The publications closes with a summary of results and the authors' considerations.

---

[6] For example, average impact per publication of 10% above the world average could correspond to below-average productivity if the average product per researcher in the field examined was less that 90.9% of the world total.

## 2. Data and method

The field of observation is all publications in the hard sciences[7] by Italian U&PROs indexed in the Thomson Reuters Web of Science (WoS), from 2001 to 2006. Observation is limited to the hard sciences because, in these, publication is a good proxy of the entire scientific output (Moed et al., 2004). The raw data acquired from Thomson Reuters were elaborated through the identification and reconciliation of all U&PROs indicated as addresses for Italian authors: 78 universities, 75 research institutions and 192 hospitals and health care research organizations were identified, using software with over 30,000 rules for matching the Italian U&PRO addresses in the 2001-2006 WoS records[8]. Observation includes articles, review articles and conference proceedings, for a total of over 250,000 publications. The relative citations observed for the publications are used as proxy of impact (Moed, 2005). Observations are made as of 30/06/2009, which is sufficient time from date of publication for confidence in the robustness of the findings on relative impact.

Publications are classified by year and by field, corresponding to the WoS subject category for the journal of publication. The citations are considered in relation to two benchmark values reported by Thomson Reuters:
- Expected Citation Rate (XCR), the average number of citations received by all world publications listed by the WoS for the same year and field[9];
- Journal Expected Citation Rate (JXCR), the average number of citations received by all publications printed in the same year in the same journal.

Standardization for XCR permits understanding of relative impact at world level. Standardization for JXCR is conducted only for publications in "top" journals, meaning journals with impact factor[10] in the top decile for distribution in each field. For those specific publications in high prestige journals, this permits understanding of relative impact when compared to other publications of equal potential.

Analysis of average impact of Italian research compared to the world average is based on aggregation of publications at a variety of levels of analysis, with calculation of the average of field-standardized impact of each publication.

## 3. Impact of Italian research compared to world average

As an example of "world ranking", distorted by use of bibliometric indicators without field-standardization, we might suggest consultation of the site for SCImago Country Rankings[11]. Here, for 2001-2006, Italy works out at eighth place in rankings for number of publications and seventh for total world citations. Limiting comparison to the top 20 nations by number of publications, Italy rates eighth for average citations per publications and ninth excluding self-citations. Naturally, these rankings are not based on field-standardized analysis and result of little interest.

---

[7] Mathematics; Physics; Chemistry; Earth and space sciences; Biology; Biomedical research; Clinical medicine; Engineering.
[8] The case of the University of Rome "Tor Vergata" offers an example of the need for and complexity of these elaborations: the authors identified 150 different bibliometric addresses for this institution over the six years examined.
[9] Publications in multidisciplinary journals are assigned to all fields associated with the relative journals and standardization is carried out with respect to the average of the XCRs for all the individual fields.
[10] Extracted from the Thomson Reuters *Journal Citation Report*, 2008.

Turning to the data for the current study, Table 1 shows a time series for all Italian publications per document type (2001-2006), with a total of 250,000 records. The general trend is for continuous increase in order. Over the six years, publications grow 26%, due particularly to an increase in articles, which are over two thirds of total publications. Growth in reviews is also very substantial. In 2006 these arrive at over 5% of total. The data on conference proceedings are anomalous: numbers only increase over 2002-2003, remaining flat over the next biennium and actually retreating in the last year. It must be noted that the increase of Italian publications is not due to a relevant increase in the number of Italian journals indexed in WoS. Italian journals indexed in 2008 Thomson Reuters' Journal Citation Report (JCR) are 82, accounting roughly for 1% of JCR journals.

| Year | Articles | Proceedings | Reviews | Total |
|---|---|---|---|---|
| 2001 | 25,956 (69.5%) | 10,195 (27.3%) | 1,202 (3.2%) | 37,353 |
| 2002 | 26,785 (70.0%) | 10,160 (26.5%) | 1,337 (3.5%) | 38,282 |
| 2003 | 28,090 (67.1%) | 12,330 (29.4%) | 1,449 (3.5%) | 41,869 |
| 2004 | 29,638 (67.9%) | 12,305 (28.2%) | 1,726 (4.0%) | 43,669 |
| 2005 | 30,904 (67.9%) | 12,643 (27.8%) | 1,960 (4.3%) | 45,507 |
| 2006 | 32,662 (69.3%) | 12,044 (25.5%) | 2,458 (5.2%) | 47,164 |
| Total | 174,035 (68.6%) | 69,677 (27.4%) | 10,132 (4.0%) | 253,844 |

*Table 1: Time series of Italian publications per document type (hard sciences; source WoS)*

We now present the analysis of field-standardized average impact per Italian publication, compared to world average: first at the overall level, then by discipline and by field. Then we continue the analysis at the organizational level, by individual institution and by organizational unit within the institution.

Table 2 presents the time series at the overall level of all U&PROs, for the two defined indicators. We remind the reader that citations to all publications are counted on June 30, 2009. The value for overall standardized impact (Cites/XCR) is consistently greater than one. This means that Italian publications are on average cited more than the world average (12% more over the six year period). The trend is clearly for constant growth, with a peak in 2006, when Italian publications were cited 30% more than the world average.

|  | 2001 | 2002 | 2003 | 2004 | 2005 | 2006 | Average |
|---|---|---|---|---|---|---|---|
| Cites/XCR | 1.00 | 1.07 | 1.03 | 1.09 | 1.18 | 1.30 | 1.12 |
| Publications in top journals (%) | 9.7 | 10.0 | 9.7 | 10.1 | 10.9 | 10.5 | 10.2 |
| Cites/JXCR | 1.09 | 1.09 | 1.05 | 1.09 | 1.10 | 1.13 | 1.09 |

*Table 2: Time series of standardized average scientific impact for the Italian research system*

The percentage of publications in top journals is also clearly increasing. Such publications receive 9% more citations than the average of citations for publications in the same year and same journal (last column, Table 2). Year 2003 shows particularly low values for the three indicators, which might be related to the peak, in the same year, for conference proceedings (Table 1). Proceedings average fewer citations than both articles and reviews.

---

[11] SCImago country rankings. http://www.scimagoir.com/ (last accessed 18 February 2011).

In terms of field-standardized average impact per publication, Italy thus places above the world average. The next sections provide detail on single disciplines and fields.

### 3.1 Field-standardized average impact per discipline and field

#### 3.1.1 Impact per discipline

Italian research product is not especially concentrated by discipline; however disciplines with more publications include clinical medicine, engineering and physics. These three disciplines account, roughly equally, for 57% of total publications (Table 3). The best performance for standardized average impact is in chemistry and clinical medicine, where publications received 19% and 20% more than the world average field-standardized citations. After these two disciplines the performances descend, for biomedical research (10% higher than world average), physics (8%), mathematics (7%), and earth and space (+2%). The impact for biology is exactly equal to the world average. Engineering is the only discipline that registers lower than world average.

Table 3 also presents incidence for publications in "top" impact factor journals. Clinical medicine (15.5%), chemistry (13.1%), this time with biomedical research (13.0%), stand out again. For all other disciplines the related percentages are under 8%. In the analysis of relative impact of publications in top journals, Chemistry publications are notable for a low value of the Cites/JXCR ratio, at 0.99, meaning that these receive fewer citations than other world publications with the same potential. At the opposite end, high values are seen in the areas of clinical medicine and mathematics (both 1.14), followed by biomedical research (1.11) and Biology (1.10).

Time-series analysis shows that over the six years, the standardized average impact increases significantly in all eight disciplines (Table 4).

|  | All journals |  | Top journals |  |
|---|---|---|---|---|
| Discipline | # of publications | Cites/XCR | # of publications (% on total) | Cites/JXCR |
| Biology | 41,846 (13.1%) | 1.00 | 7.5 | 1.10 |
| Biomedical research | 39,957 (12.5%) | 1.10 | 13.1 | 1.11 |
| Chemistry | 26,987 (8.4%) | 1.19 | 13.0 | 0.99 |
| Clinical medicine | 61,541 (19.2%) | 1.18 | 15.5 | 1.14 |
| Earth and space sciences | 16,671 (5.2%) | 1.02 | 6.8 | 1.05 |
| Engineering | 60,583 (18.9%) | 0.94 | 4.8 | 1.07 |
| Mathematics | 13,296 (4.2%) | 1.07 | 6.0 | 1.14 |
| Physics | 59,499 (18.6%) | 1.08 | 4.7 | 1.03 |
| Total publications/Average | 253,844* | 1.12 | 25,857 (10.2) | 1.09 |

*Table 3: Standardized average impact of Italian publications per discipline, data 2001-2006*
*\* This figure is lower than the column total because of double counting of single publications in multiple disciplines.*

| Discipline | 2001 | 2002 | 2003 | 2004 | 2005 | 2006 | Average |
|---|---|---|---|---|---|---|---|
| Biology | 0.87 | 0.97 | 0.93 | 0.96 | 1.02 | 1.15 | 1.00 |
| Biomedical research | 0.94 | 1.05 | 0.98 | 1.06 | 1.20 | 1.33 | 1.10 |
| Chemistry | 1.10 | 1.09 | 1.12 | 1.17 | 1.27 | 1.38 | 1.19 |
| Clinical medicine | 1.01 | 1.11 | 1.06 | 1.18 | 1.30 | 1.38 | 1.18 |
| Earth and space sciences | 0.86 | 0.93 | 0.94 | 1.02 | 1.12 | 1.15 | 1.02 |
| Engineering | 0.83 | 0.83 | 0.87 | 0.87 | 1.03 | 1.17 | 0.94 |
| Mathematics | 0.97 | 1.05 | 1.03 | 1.03 | 1.05 | 1.24 | 1.07 |
| Physics | 1.04 | 1.14 | 0.97 | 1.04 | 1.09 | 1.21 | 1.08 |
| Average | 1.00 | 1.07 | 1.03 | 1.09 | 1.18 | 1.30 | 1.12 |

*Table 4: Time series of standardized average scientific impact (Cites/XCR) per discipline, data 2001-2006*

In 2006, the last year of the series, the ratio of citations received to expected is never less than the 1.15 seen for earth sciences, biology and space sciences, and reaches a maximum of 1.38 in chemistry and clinical medicine. Time-series analysis was also conducted for Cites/JXCR, but we do not present it here, for reasons of space.

Table 5 presents the average annual variations in the indicators of both Cites/JXCR and Cites/XCR. Over the six years, the average value of the Cites/XCR ratio increases at an average rate of 5.9% (second column, last row). The disciplines with greatest average annual increase (+7.4%) are engineering and biomedical research, and the one with least increase (+3.6%) is physics. Presence in top journals also increases, at an overall average annual rate of +3.2%. However, performance in individual disciplines is quite differentiated, with some fields strengthening presence in top journals (especially mathematics and physics), while other have average annual growth of almost nil or even negative (engineering, -0.1%; clinical medicine, -1.8%). The average increase in Cites/JXCR ratio is less notable (+0.5%) and the distribution for the different disciplines is quite mixed. The greatest average increase is in earth and space sciences (+2.5%), while mathematics shows a notable decrease (-2.6%).

| Discipline | Average annual increase in Cites/XCR (%) | Average annual increase in % publications in top journals | Average annual increase in Cites/JXCR (%) |
|---|---|---|---|
| Biology | 5.7 | 3.9 | 1.3 |
| Biomedical Research | 7.4 | 3.8 | 1.4 |
| Chemistry | 4.7 | 4.1 | 0.4 |
| Clinical Medicine | 6.7 | -1.8 | 1.2 |
| Earth and Space Sciences | 6.2 | 3.7 | 2.5 |
| Engineering | 7.4 | -0.1 | 0.2 |
| Mathematics | 5.4 | 6.5 | -2.6 |
| Physics | 3.6 | 5.6 | -0.1 |
| Average | 5.9 | 3.2 | 0.5 |

*Table 5: Average annual variation in standardized average impact indicators per discipline, data 2001-2006*

### 3.1.2 Impact per field

Within each discipline it is possible to conduct deeper analysis at the level of individual fields. We present the example of the fields in the biomedical research discipline.

Within biomedical research, the field of medical laboratory technology is relatively small for output but achieves a received to expected citations ratio of 1.44, which puts it at the top for standardized average impact (Cites/XCR), followed by hematology (1.31) and allergy (1.22) (Table 6). Although the publications for the entire discipline receive 10% more citations than the world average, there are four fields that get less than average, with the field of anatomy and morphology dipping to 29% below average. The Cites/XCR ratio increased in all 14 fields over the six years: maximum increase was for infectious diseases (+14.5%), followed by virology (+11.5%) and medical laboratory technology (+11.4%).

Still referring to Table 6, we can see that hematology and medicinal chemistry show significant concentration of publications in top journals (28% and 22.1% of the total of Italian publications for these fields). Notable increase for presence in top journals is seen in toxicology (+32.8% average rate of annual increase) and medical laboratory technology (+31.2%). As seen previously, Italian biomedical research published in top journals receives an average of 11% of citations more than the average of all works published in the same journals (last column, last line). Four fields exceed this 11% average for Cites/JXCR. Overall, the single field values for this ratio vary substantially, from a minimum of 0.9 for virology to a maximum of 1.21 for hematology, but the distribution still appears flat compared to that for CITES/XCR (column 2).

| Field | Cites/XCR | | Publications in top journals | | Cites/JXCR (average 2001-06) |
| --- | --- | --- | --- | --- | --- |
| | Average 2001-06 | Average annual increase % | Average 2001-06 (%) | Average annual increase % | |
| Medical laboratory technology | 1.44 | 11.4 | 16.1 | 31.2 | 1.08 |
| Hematology | 1.31 | 6.5 | 28.0 | -0.6 | 1.21 |
| Allergy | 1.22 | 10.5 | 12.7 | 15.3 | 1.13 |
| Chemistry, medicinal | 1.20 | 9.1 | 22.1 | 8.5 | 0.97 |
| Immunology | 1.14 | 6.2 | 8.1 | 4.6 | 1.12 |
| Medicine, research & experimental | 1.06 | 4.0 | 10.5 | 4.9 | 1.07 |
| Pharmacology & pharmacy | 1.06 | 6.4 | 8.6 | -3.4 | 1.04 |
| Toxicology | 1.05 | 5.0 | 5.7 | 32.8 | 1.04 |
| Oncology | 1.04 | 9.6 | 11.1 | 5.8 | 1.18 |
| Pathology | 1.03 | 5.6 | 10.1 | 0.7 | 0.94 |
| Virology | 0.94 | 11.5 | 0.3 | 0.8 | 0.90 |
| Infectious diseases | 0.93 | 14.5 | 13.7 | 7.2 | 0.95 |
| Radiology, nuclear medicine & medical imaging | 0.79 | 10.4 | 10.2 | 12.1 | 1.09 |
| Anatomy & morphology | 0.71 | 4.9 | 0.1 | 2 | 0.92 |
| Average | 1.10 | 7.4 | 13.1 | 3.2 | 1.11 |

*Table 6: Standardized average impact of Italian publications per individual fields of biomedical research, data 2001-2006*

### 3.2 Field-standardized average impact at organizational level

The analyses of the previous sections show some strong and weak points of the Italian scientific system, the trends at general and single discipline levels, and the case for the single fields of one selected discipline. Analysis of the average impact of national research, in order to serve in policy formation, should also provide information at the institutional level.

**3.2.1 Impact per type of institution**

There are three types of organizations in the Italian public research system: universities, which produce over 67% of total research output, research institutions, and hospitals-HCROs, which respectively provide 21% and 11% of total output (Table 7, last line). The indices of concentration for each discipline clearly show the general specializations by type of organization[12]. Hospitals and healthcare research organizations concentrate on two disciplines: biomedical research and clinical medicine, with their concentration of production reaching 29.5% and 27.4% of the national totals. Research institutions are particularly active in physics and earth and space sciences, where they achieve shares of 40.7% and 35.3% of total Italian research product. Universities are active on all disciplines but hold an almost complete monopoly on mathematics, where they achieved 90% of total national production for 2001-2006.

| Discipline | Universities | Research institutions | Hospitals and HCROs |
|---|---|---|---|
| Biology | 70.0 (1.03) | 20.8 (0.98) | 9.2 (0.83) |
| Biomedical research | 61.4 (0.9) | 9.1 (0.43) | 29.5 (2.66) |
| Chemistry | 75.9 (1.12) | 22.9 (1.08) | 1.2 (0.11) |
| Clinical medicine | 65.5 (0.96) | 7.1 (0.33) | 27.4 (2.47) |
| Earth and space sciences | 62.7 (0.92) | 35.3 (1.67) | 2.0 (0.18) |
| Engineering | 76.7 (1.13) | 21.5 (1.01) | 1.7 (0.15) |
| Mathematics | 89.2 (1.31) | 10.8 (0.51) | 0.0 (0.0) |
| Physics | 58.8 (0.87) | 40.7 (1.92) | 0.5 (0.05) |
| Average | 67.9 | 21.2 | 11.1 |

*Table 7: Division of research output by type of institution per each discipline in Italy (concentration indices in brackets), data 2001-2006*

Of the three types of organizations, hospitals-HCROs produce publications that record the greatest average impact, with 20% more citations than the world average, compared to 16% for publications from research institutions and 8% for those from universities (Table 8, column 2). Hospitals-HCROs are also first for presence in top journals: 15.1% of the total of publications from these organizations go to top journals, compared to 9.2% from universities and 7.6% from research institutions. Hospital and HCROs also dominate results for the last indicator considered (Cites/JXCR), with higher performance (1.16) than the other two types of organizations (1.06).

| Organization | Cites/XCR | Publications in top journals (%) | Cites/JXCR |
|---|---|---|---|
| Universities | 1.08 | 9.2 | 1.06 |
| Research institutions | 1.16 | 7.6 | 1.06 |
| Hospitals and HCROs | 1.20 | 15.1 | 1.16 |

*Table 8: Standardized average impact of Italian publications per type of institution, data 2001-2006*

In summary, although university research is much greater in quantity, it seems that other types of organizations achieve greater results in terms of average impact. Hospitals and HCROs produce work that has a particularly notable average level of impact. This is a numerous and geographically scattered group of institutions, numbering almost 200 in total, but evidently particularly focused on biomedical research and clinical medicine, fields in which they produce works of decidedly higher average impact.

---

[12] Concentration indices shown in brackets in Table 7 represent a measure of association between two variables based on frequency data, which varies around the neutral value of 1. For example, in biology, the value of 1.03 for universities derives from the ratio of two ratios: ratio of total universities' publications in biology to all Italian publications in biology (70.0) divided by ratio of total universities' publications to all Italian publications (67.9).

### 3.2.2 Impact per individual organization

Two thirds of the 2001-2006 research output is concentrated in 15 of the nation's organizations (Table 9), representing three research institutions and 12 universities. Of the 15 organizations, the National Institute for Astrophysics (INAF), has the highest average ratio of citations received to expected (1.36), followed by the universities of Turin (1.29), Pavia (1.28) and Milan (1.26). These three universities lead the rankings for presence in top journals: 13.3% of works produced by researchers at the University of Pavia are published in top journals, while the universities of Milan and Turin achieve levels of 13.1% and 12.5%. Examination of Column 5 finds two physics institutions with lackluster performance in share, with top journal publication of only 3.2% and 5.3% of works. However the data in Column 6 show that these works are truly excellent, since they receive an average of 29% (for the INAF) and 15% (for the INFN) more citations than other publications in the same journals.

| Organization | Type* | All journals | | Top journals | |
|---|---|---|---|---|---|
| | | # of public. | Cites/XCR | # of public. (% on total) | Cites/JXCR |
| National Research Council | RI | 33,490 | 1.22 | 11.7 | 1.02 |
| University of Rome "Sapienza" | U | 17,967 | 1.09 | 10.7 | 0.99 |
| University of Bologna | U | 14,246 | 1.24 | 10.6 | 1.14 |
| University of Milan | U | 14,112 | 1.26 | 13.1 | 1.08 |
| University of Padua | U | 13,139 | 1.24 | 11.2 | 1.04 |
| University of Naples "Federico II" | U | 12,240 | 1.09 | 10.7 | 0.93 |
| National Institute of Nuclear Physics (INFN) | RI | 10,630 | 1.23 | 5.3 | 1.15 |
| University of Pisa | U | 10,326 | 1.15 | 8.7 | 1.06 |
| University of Florence | U | 10,297 | 1.19 | 12.1 | 1.10 |
| University of Turin | U | 8,542 | 1.29 | 12.5 | 1.13 |
| University of Genoa | U | 7,724 | 1.09 | 10.2 | 1.09 |
| University of Rome "Tor Vergata" | U | 7,172 | 1.15 | 10.3 | 0.98 |
| National Institute of Astrophysics (INAF) | RI | 6,533 | 1.36 | 3.2 | 1.29 |
| University of Pavia | U | 6,170 | 1.28 | 13.3 | 1.04 |
| University of Bari | U | 5,982 | 1.10 | 10.6 | 1.03 |

*Table 9: Standardized average impact of publications per Italian organization; data 2001-2006, limited to the top 15 organizations for number of publications*
 * U = university; RI = research institution

Table 10 presents a list of the top 10 organizations for Cites/XCR, from among those organizations that achieved at least 50 publications over the six years observed[13]. There are no universities in the list. Four of the organizations are research institutes and six are hospitals-HCROs. The Italian Space Agency tops the list: its 137 publications achieve an average Cites/XCR value of 2.91. After this come the Casa di Cura Columbus (2.89) and the S. Luigi Gonzaga Hospital (2.51). The placement of publications by the Italian Space Agency is not particularly admirable: only 6 (4.4%) are published in top journals. This means that the high ranking of this organization is mainly due to the high number of citations received by articles published in journals other than top. At the other extreme of the group are the Casa di Cura Columbus and the A. Buzzati European Laboratory for Molecular Biology, where almost half of the total works appear in high-impact journals. However, the value of Cites/JXCR (0.84) is very low for the latter organization, which means that although its works are published in top journals, they receive fewer citations than works by other organizations published in the same journals.

---
[13] It should be noted that nine out of ten ranked institutions have quite few publications. As a consequence the extract of publications published in top journal is very little. The resulting ranking for Cites/JXCR may be due then to one or very few articles, whose citations may determine the overall ranking. While it may be of interest to a decision maker knowing the top research institutions regardless their size, potential bias may be avoided simply increasing the minimum threshold of published articles.

| Organization | Type* | All journals | | Top Journals | |
|---|---|---|---|---|---|
| | | # of public. | Cites/XCR | # of public. (% on total) | Cites/JXCR |
| Italian Space Agency | RI | 137 | 2.91 | 4.4 | 1.93 |
| Casa di Cura Columbus | H | 128 | 2.89 | 46.9 | 1.51 |
| S. Luigi Gonzaga Hospital | H | 143 | 2.51 | 23.1 | 2.14 |
| A. Buzzati Europ. Lab. of Molecular Biology | RI | 98 | 2.38 | 46.9 | 0.84 |
| Busto Arsizio Civic Hospital | H | 60 | 2.30 | 18.3 | 2.96 |
| Vimercate Hospital | H | 61 | 2.29 | 19.7 | 3.23 |
| Inter-univ. Biotechnology Consortium (CIB) | RI | 63 | 2.19 | 20.6 | 1.16 |
| Alpine Ecology Centre | RI | 67 | 2.17 | 44.8 | 1.30 |
| Paternò Civic Hospital | H | 106 | 2.14 | 15.1 | 1.43 |
| European Institute of Oncology (HCRO) | H | 1,148 | 1.98 | 22.0 | 1.33 |

*Table 10: Standardized average impact of publications per Italian organization; data 2001-2006, limited to the top 10 organizations for CITES/XCR from those with a minimum of 50 total publications*
*\* H = hospitals and health care research organizations; RI = research institution*

Ranking lists can be formulated for other impact indicators as well. Furthermore, analyses can be carried out at discipline and field levels. Finally, the measurement of the average impact for an organization's full scientific portfolio can be integrated with further analysis focused on a limited set of total scientific production from each organization, for example the top decile of the organizations' publications, as rated for field-standardized impact. Comparative evaluation would then permit qualification of the level of excellence of an organization relative to its cutting edge scientific production. Few examples of this, which are of interest to the country-specific decision maker are presented in the appendix. Findings confirm what emerged in the preceding sections: in the Italian public system, research conducted at hospitals and HCROs produces scientific results with higher average impact than for other organizations, with the specific disciplines of clinical medicine and biomedical actually performing better than for all other disciplines.

### 3.2.3 Impact of research units within individual institutions

U&PRO administrators can use the analytical process illustrated here in various ways. The ratings for national research organizations, for the various dimensions shown, offer a benchmark system that reveals strong and weak points of each organization. The focus on single disciplines and fields offers detailed information for targeted interventions. The time-series analysis serves to detect trends, while inter-temporal analysis is apt to evaluate effectiveness of interventions. In highly complex institutions the analysis can be detailed by single research groups, to evidence and reward best practices, or to offer them as models for other organizational units. As an example, we present the National Research Council, Italy's premier public research institution, with a research staff of around 7,000, subdivided in over 100 institutes throughout the nation. The analysis is only possible due to the reconciliation of all different names indicated as the authors' "home" institution, in elaborating the initial database. Table 11 presents the findings for the top ten CNR institutes, identified for Cites/XCR. The top ten CNR institutes range in production from 120 to a maximum of 436 publications for the six-year period, working in various disciplines. Three have a Cites/XCR value of greater than 2. The ITAE (Institute for Advanced Energy Technology) clearly places above all others, with a ratio of citations received to expected of 2.92. The next two institutes are "IBAF", with Cites/XCR of 2.18 and "IMATI", at 2.12. The ITAE also places in top position for Cites/JXCR (1.96). There is actually a strong correlation between the Cites/XCR and Cites/JXCR rankings, as seen from columns 3 and 5. The Institute for Biomedicine and Immunology places at the top for concentration of works published in top journals. Almost 40% of its 416 publications were published in top-impact journals. Next for this ranking are the "IBAF" institute (32.5%) and once again the ITAE (27.5%).

This type of analysis can be detailed at the sectorial and inter-temporal levels according to the specific needs of the decision-maker.

| Organizations | All journals | | Top journals | |
|---|---|---|---|---|
| | # of public. | Cites/XCR | # of public. (% on total) | Cites/JXCR |
| CNR-ITAE: Institute for Advanced Energy Technology | 131 | 2.92 | 27.5 | 1.96 |
| CNR-IBAF: Institute for Agro-environmental and Forestry Biology | 120 | 2.18 | 32.5 | 1.85 |
| CNR-IMATI: Institute for Applied Mathematics and Information Technology | 246 | 2.12 | 15.0 | 1.54 |
| CNR-IBIM: Institute for Biomedicine and Immunology | 268 | 1.89 | 39.9 | 1.42 |
| CNR-ISTI: Institute for Information Science and Technology | 416 | 1.86 | 4.6 | 1.35 |
| CNR-ISC: Complex Systems Institute | 217 | 1.84 | 14.3 | 1.20 |
| CNR-ISTEC: Institute for Ceramic Materials Science and Technology | 291 | 1.79 | 23.7 | 1.44 |
| CNR-IIT: Institute for Telecommunications and Informatics | 150 | 1.74 | 6.7 | 1.11 |
| CNR-IMCB: Institutions for Composite and Biomedical Materials | 210 | 1.64 | 14.3 | 0.85 |
| CNR-ISMN: Institute for Study of Nano-materials | 436 | 1.61 | 14.7 | 0.98 |

*Table 11: Standardized average impact of publications for CNR institutes; data 2001-2006, limited to the top 10 organizations for CITES/XCR*

**4. Discussion and conclusions**

This work presents time-series and cross-field analysis of the Italian public research system, particularly the field-standardized average impact of research output compared to the world average, for the period 2001-2006.

The aggregate data show evident growth in national scientific production, achieving a rate of almost 5% per year over the six years observed. A more interesting and significant observation is the field-standardized average impact of Italian publications, compared to the world average, The Italian research system shows good overall performance and a positive trend. Publications by Italian researchers receive 12% more citations than the world average and this data have been in constant increase, reaching a peak (+30%) in the final year examined. The representation of works in top journals is also in clear increase, likewise the average impact of these "top-journal" works. Three disciplines seem to be the motive force behind the improving general national performance: clinical medicine, biomedical research and chemistry. In these disciplines, the standardized average impact and the percentage of works in top journals are significantly higher than in others. Universities produced over two thirds of total national research product, but it is the hospitals and health research organizations that lead for impact. This numerous group (almost 200 organizations) is focused primarily on biomedical research and clinical medicine.

The objective of the work is also to illustrate the essentials of a methodology that provides diagnostic support even at highly detailed levels. Examples were presented for methods and results concerning performance of single organizations, disciplines and detailed fields.

The methodology, until now only applied in Italy, is open to general use and it essentially replicable in any country. This methodology can support policy interventions to consolidate excellence and reinforce weak but strategic fields for national scientific development. The inter-temporal aspects of the analysis can also provide indications for the effectiveness of national interventions attempted. The same considerations apply to the case of administration for individual universities and public research organizations.

National policy interventions or related changes in single organizations cannot be considered without other dimensions of evaluation. In addition to producing research, U&PROs are also responsible for transferring results to the productive system, and universities bear the crucial responsibility of teaching. Even considering research alone, measurement of effectiveness should not inform policy formulation without joint consideration of efficiency. In this study we have proposed indicators of average impact of research output, but the related labor and capital inputs should also be the subject to comparative evaluation. There is objective difficulty in world-level comparative measurement of productivity. The authors have succeeded in such analysis at the domestic level (Abramo and D'Angelo, 2011), and are able to compare field-standardized research productivity both at the individual and organizational level. To carry out international comparisons of research productivity it is necessary that other nations as well provide such measures.

**APPENDIX – further analysis**

Table 12 presents the list of the top 10 organizations for percentage of publications in top journals, from those with at least 50 publications over the 2001-2006 period. We see that all these organizations but one are hospitals, and that in all cases but one their publications receive average citations that are higher than for other works in the same top journals.

|  |  | All journals | | Top Journals | |
|---|---|---|---|---|---|
| Organization | Type* | # of public. | Cites/XCR | # of public. (% on total) | Cites/JXCR |
| Casa di Cura Columbus | H | 128 | 2.89 | 46.9 | 1.51 |
| A. Buzzati Europ. Lab. of Molecular Biology | H | 98 | 2.38 | 46.9 | 0.84 |
| Alpine Ecology Centre | RI | 67 | 2.17 | 44.8 | 1.30 |
| "Bianchi e Melacrino Morelli" Hospital | H | 210 | 1.69 | 33.8 | 1.38 |
| "Riuniti" Hospital of Bergamo | H | 768 | 1.78 | 32.9 | 1.29 |
| San Carlo Borromeo Hospital | H | 133 | 1.68 | 31.6 | 1.28 |
| G.B. Bietti Found. for Research in Ophthalmology (HCRO) | H | 61 | 1.19 | 31.1 | 1.05 |
| Valduce Hospital | H | 55 | 1.39 | 30.9 | 1.35 |
| V. Cervello Hospital | H | 142 | 1.95 | 28.9 | 1.50 |
| Lecco Hospital | H | 191 | 1.80 | 28.8 | 1.28 |

*Table 12: Standardized average impact of publications per Italian organization; data 2001-2006, limited to the top 10 for incidence of "top journal" publications from those organizations with a minimum of 50 total publications*

* H = hospitals and health care research organizations; RI = research institution.

Table 13 presents the example of a list of the top ten organizations for standardized average impact (Cites/XCR), for the top 10% of their publications. Seven of these organizations are also present in Table 10 and the exact same three organizations hold the top three places in both tables.

|  |  |  | Cites/XCR (average) | |
|---|---|---|---|---|
| Organization | Type* | # of public. | All publications | Top 10% |
| Italian Space Agency | RI | 137 | 2.91 | 18.99 |
| Casa di Cura Columbus | H | 128 | 2.89 | 15.79 |
| S. Luigi Gonzaga Hospital | H | 143 | 2.51 | 15.22 |
| Vimercate Hospital | H | 61 | 2.29 | 14.75 |
| Busto Arsizio Civic Hospital | H | 60 | 2.30 | 12.20 |
| "S. Croce e Carle" Hospital | H | 206 | 1.84 | 11.00 |
| V. Cervello Hospital | H | 142 | 1.95 | 9.87 |
| European Institute of Oncology (HCRO) | H | 1,148 | 1.98 | 9.36 |
| Lecco Hospital | H | 191 | 1.80 | 9.32 |
| Inter-university Consortium for Biotechnology | RI | 63 | 2.19 | 9.11 |

*Table 13: Standardized average impact of publications per Italian organization; data 2001-2006, limited to the top 10 for average CITES/XCR of their top 10% publications, from those organizations with at least 50 total publications*

* H = hospitals and health care research organizations; RI = research institution.

The study of organizations through the impact of their total research output risks hiding differences concerning their internal disciplines and fields. The analysis can be detailed to reveal this level of data. Table 14 presents information for the ten best national organizations in the physics discipline, identified for Cites/XCR. The top three organizations are research institutes, which all score above two for Cites/XCR. There are also four universities, two of which are schools for advanced studies (Pisa School for Advanced Studies; Trieste International School for Advanced Studies).

| Organization | Type* | All journals | | Top journals | |
|---|---|---|---|---|---|
| | | # of public. | Cites/XCR | # of public. (% on total) | Cites/JXCR |
| Italian Space Agency | RI | 108 | 3.59 | 5.6 | 1.93 |
| Istituto Superiore di Sanità | RI | 167 | 2.23 | 14.4 | 1.43 |
| National Institute for Geophysics and Volcanology | RI | 92 | 2.01 | 8.7 | 1.21 |
| Pisa School for Advanced Studies | U | 874 | 1.83 | 7.9 | 1.19 |
| European Centre for Theoretical Studies in Nuclear Physics | RI | 187 | 1.77 | 5.9 | 0.89 |
| Trieste International School for Advanced Studies | U | 1,436 | 1.74 | 8.9 | 1.10 |
| Europ. Laboratory for Non-linear Spectroscopy | RI | 220 | 1.59 | 19.5 | 0.77 |
| University of Insubria | U | 606 | 1.58 | 11.9 | 1.01 |
| University of Eastern Piedmont "A. Avogadro" | U | 210 | 1.54 | 1.4 | 2.08 |

*Table 14: Standardized average impact of publications in physics per Italian organization; data 2001-2006, limited to the top 10 organizations for Cites/XCR, from those with a total of at least 50*
 * U = university; RI = research institution.

The analysis can inquire deeper, for example to the level of fields. Table 15 presents the list of the top ten national organizations for research production in oncology, as identified for Cites/XCR.

| Organization | Type* | All journals | | Top journals | |
|---|---|---|---|---|---|
| | | # of public. | Cites/XCR | # of public. (% on total) | Cites/JXCR |
| Vita-Salute San Raffaele University | U | 64 | 2.42 | 29.7 | 1.84 |
| Paternò Civic Hospital | H | 68 | 2.07 | 8.8 | 1.52 |
| University of Ferrara | U | 136 | 2.01 | 16.9 | 1.43 |
| Bellaria Hospital | H | 76 | 1.95 | 14.5 | 2.32 |
| Perugia Hospital | H | 97 | 1.73 | 14.4 | 2.20 |
| San Raffaele (HCRO) | H | 186 | 1.72 | 30.1 | 1.17 |
| Humanitas (HCRO) | H | 75 | 1.60 | 14.7 | 1.24 |
| University of Verona | U | 183 | 1.52 | 12.6 | 1.20 |
| Carlo Besta Neurological Institute (HCRO) | H | 59 | 1.51 | 8.5 | 2.43 |
| University of Sassari | U | 84 | 1.51 | 14.3 | 1.44 |

*Table 15: Standardized average impact of publications in oncology per Italian research organizations; data 2001-2006, limited to the top 10 organizations for CITES/XCR, from those with a total of at least 50 publications*
 * H = hospitals and health care research organizations; U = university